\documentclass{TTP_DSL2006}

\usepackage[dvipsone]{graphicx}
\usepackage[intlimits]{amsmath}
\usepackage{amssymb}
\usepackage{exscale}
\usepackage[colorlinks,urlcolor=blue]{hyperref}
\begin{document}

\title{ Materials Bound by Non-Chemical Forces: External Fields and the Quantum Vacuum} 

\author{ John Swain \inst{1}$^{,\rm{a}}$ \textbf{,} Allan Widom \inst{1}$^{,\rm{b}}$ and Yogendra N. Srivastava \inst{1,2}$^{,\rm{c}}$}

\institute{Northeastern University, Department of Physics, 110 Forsyth Street, Boston, MA, 02115, USA
\and
Physics Department, Universit\'a di Perugia \& INFN, Perugia, Italy}
\maketitle

\vspace{-3mm}
\sffamily
\begin{center}
$^{a}$john.swain@cern.ch, $^{ b}$allan.widom@gmail.com, $^{c}$yogendra.srivastava@gmail.com
\end{center}

\vspace{2mm} \hspace{-7.7mm} \normalsize \textbf{Keywords:} materials, phase diagram, floating water bridge,
ferrofluids,Casimir effect, quantum vacuum, red blood cell, erythrocyte, self-assembly, nanofabrication, negative pressure\\

\vspace{-2mm} \hspace{-7.7mm}
\rmfamily
\noindent \textbf{Abstract.} 
We discuss materials which owe their stability 
to external fields. These include: 1) external electric or magnetic fields, and 2) quantum vacuum fluctuations in
these fields induced by suitable boundary conditions (the Casimir effect). Instances of the first case include
the floating water bridge and ferrofluids in magnetic fields. An example of the second case is taken from 
biology where the Casimir effect provides an explanation of the formation of stacked aggregations or ``rouleaux'' by negatively
charged red blood cells. We show how the interplay between electrical and Casimir forces can be used to drive self-assembly of
nano-structured materials, and could be generalized both as a probe of Casimir forces and as a means
of manufacturing nanoscale structures. Interestingly, all the cases discussed involve the generation of the
somewhat exotic negative pressures. We note that very little is known about the
phase diagrams of most materials in the presence of external fields other than those represented by the macroscopic
scalar quantities of pressure and temperature. Many new and unusual states of matter may yet be undiscovered.

\vspace{-2mm}
\section{Introduction}
\vspace {-3mm}

\hspace{4.5mm} It is common knowledge that the mechanical properties of a material depend on
external influences, the most commonly-considered of these being pressure P and temperature T. These
can be represented as scalar fields, often uniform over the material considered.
Somewhat less attention has been paid to the effects of external fields and it is to this issue we now turn.

\section{External Electric and Magnetic Fields}

In 2007 the floating water bridge \cite{Armstrong} was rediscovered\cite{waterbridge-Elmar}, generating
a major wave of interest. The phenomenon is as follows: two containers
with pure water are filled to the brim and placed in contact with each other. A high voltage of about
20 kV is applied via electrodes placed in the water. After some preliminary arcing, a bridge of water forms
joining the two. If the beakers are pulled apart slowly, a thin rod-like piece of water - a ``floating water bridge''
- forms connecting the two. It can be a centimeter or more in length and has the appearance of a glass
rod. A piece of thread can be pulled through it, and while it will bend somewhat, the thread can pass through
without breaking the bridge. Clearly, this is water, but not as we normally know it.

The phenomenon is quite robust and can be easily replicated (with care to avoid electric shocks)
using an old-fashioned colour TV to provide 25 kV, ordinary drinking glasses, and consumer grade
distilled water from a pharmacy\cite{waterbridge-demo}.

After Nature\cite{waterbridge-Nature} dedicated a page to this mysterious phenomenon, it was shown that it in fact had quite a simple
interpretation\cite{waterbridge-us} - the water is held up by Maxwell stresses due to electric flux lines threading the water bridge.
A simple calculation shows that the tension (negative pressure) per unit cross sectional area of the water bridge with an applied electric
field $\vec{E}$ is $\left(\frac{\epsilon-1}{4\pi}\right)|\vec{E}|^2$ where $\epsilon$ is the dielectric constant of water, which at DC or very low
frequencies is
large and around 80.
There is a simple ferrofluid analog which is ferrofluid suspended between the poles of a magnet, and a simple
analog known to most school children is a thread of iron filing or chain of unlinked paperclips suspended between
the poles of a horseshoe magnet. The lesson to be learned from this is that an external electric or magnetic field can significantly alter the
mechanical properties of a material.

The water bridge remains under intensive investigation\cite{waterbridge-ongoing}. While to a first approximation the explanation
set out in \cite{waterbridge-us} can explain a great deal of the statics and dynamics of the water bridge with
water modelled as simple dielectric liquid, it remains an open question whether
or not there are distinct phases of water appearing with different applied electric fields. The question of phase
diagrams in 3 or more dimensions is largely unexplored. Of course for magnetic system, phase diagrams
are routinely drawn in the temperature-magnetic field plane, but now with pressure suppressed. What new
and interesting phases might be present in higher-dimensional phase diagrams? This would seem to be
a very open field.

There are also clear practical implications in that the mechanical strengths of materials can be changed
by external fields. Water in a floating water bridge clearly exhibits a rigidity, and indeed a negative
pressure (tension) due to an applied electric field. For another simple example consider the work required
to separate a soft iron rod into two pieces after a crack forms completely severing the material. Without
an external magnetic field very little force is needed, while if an external magnetic field is applied along
the rod, one essentially has to pull apart two facing North and South poles which attract each other -- the
iron is effectively made stronger by an external applied magnetic field.

\section{External Fields due to the Quantum Vacuum}

In classical mechanics, the vacuum is simply empty space and unaffected by external fields
or boundaries. Not so in quantum field theory, where there is a zero point energy $\frac{1}{2}\hbar\omega$
for each normal mode of frequency $\omega$. Formally this sum is divergent and usually neglected,
but the modes involved and thus the sum can be modified by boundaries. This is the basis of the
Casimir effect \cite{Casimir} in which two parallel uncharged metal plates attract each with a force
that varies as $1/d^4$ with the distance between them. This can be seen intuitively in the following
way: as the plates are moved closer together wavelengths long compared to $d$ do not contribute
to the (formally infinite) sum. Since the energy decreases as the plates are brought together, this
constitutes an attractive force, due to the restructuring of field modes in the quantum vacuum. This
force has been measured in the laboratory in 1958 \cite{Sparnaay} and again later \cite{Lamoreaux-etc}.

In macroscopic systems, the Casimir effect is usually obscured by Coulomb-type forces falling
more slowly with distance. However, as was pointed out long ago in \cite{Phys-rep}, at the nanometer scale
Casimir forces can become competitive with Coulomb ones.

An interesting example of this was studied in \cite{blood} which addressed the problem of the formation
of stacks or ``rouleaux'' of erythrocytes (red blood cells) despite their carrying a negative charge which
should make them repel each other. If one writes the total energy as 
a function of separation of the erythrocytes including a repulsive Debye-screened (considering blood as
an ionic solution - basically salty water) Coloumb energy
as well as an attractive Casimir energy between dielectric plate, then with reasonable physiological
parameters, one obtains a phase diagram with a phase in which rouleaux form and another
in which they do not. Note that the Casimir energy also corresponds to a negative pressure between the cells, pulling them
together.

Following \cite{blood}, we model erythrocytes as
dielectric plates of area $A$ with surface charge density $\sigma$ and 
dielectric constant $\epsilon_1$ in a fluid of dielectric constant $\epsilon_2$
separated by $d<<\sqrt{A}$.
With $\Lambda$ the Debye screening length and
$
\upsilon=c\left[(\varepsilon_1-\varepsilon_2)
/(\varepsilon_1+\varepsilon_2)\right]^{2}
$
the total free energy per unit area $u$ as a function of plate separation
$d$ is\cite{Phys-rep}:
\begin{equation}
u(d)=\frac{\sigma^2\Lambda}{2\varepsilon_2}
\left\{e^{-d/\Lambda}-
\left(\frac{\pi^{2}\hbar \upsilon
\sqrt{\varepsilon_0\varepsilon_2}}
{360\sigma^{2}\Lambda}\right)\frac{1}{d^{3}}\right\}.
\label{free}
\end{equation}
The first term on the right describes the Coulomb
repulsion between red blood cells while the second term
describes the Casimir attraction between red blood
cells. The relative strength of the effects can be
described by the dimensionless parameter
\begin{equation}
a=\left(\frac{\pi^{2}\hbar \upsilon \sqrt{\varepsilon_0\varepsilon_2}}
{360\sigma^2\Lambda^{4}}\right)
=\left(\frac{\pi^{2}\hbar c \sqrt{\varepsilon_0\varepsilon_2}}
{360\sigma^2\Lambda^{4}}\right)\left[\frac{(\varepsilon_1-\varepsilon_2)}
{(\varepsilon_1+\varepsilon_2)}\right]^{2}.
\label{a_parameter}
\end{equation}
The Debye screening length \begin{math} \Lambda \end{math} at temperature $T$ is related
to the ionization strength \begin{math} I \end{math} via
\begin{math} \Lambda^2 =\{\varepsilon_2 k_BT/e^2\tilde{I}\} \end{math}.
The ionization strength in physical units is
\begin{math} \tilde{I}=\sum_a z_a^2 n_a \end{math}
where \begin{math} n_a \end{math} is the number of ions/m$^3$
having an ionic charge \begin{math} z_a |e|\end{math}. (In units
of moles per liter, one employs the ionization strength
\begin{math} I=[10^{-3}{\rm meter^3/liter}](\tilde{I}/N_A) \end{math}
where Avogadro's \begin{math} N_A \end{math} is the number of ions
per mole). Fig. 2 illustrates the free energy per unit area as a function of $d/\Lambda$
showing that a minimum
only forms for sufficiently small values of $a$. The phase diagram is shown
in Fig. 3 for a reference
ionization strength $I_{0}=0.05\ $moles/liter and room temperature.

\begin{figure}[h]
\begin{center}
\begin{minipage}{12pc}
\includegraphics[width=12pc]{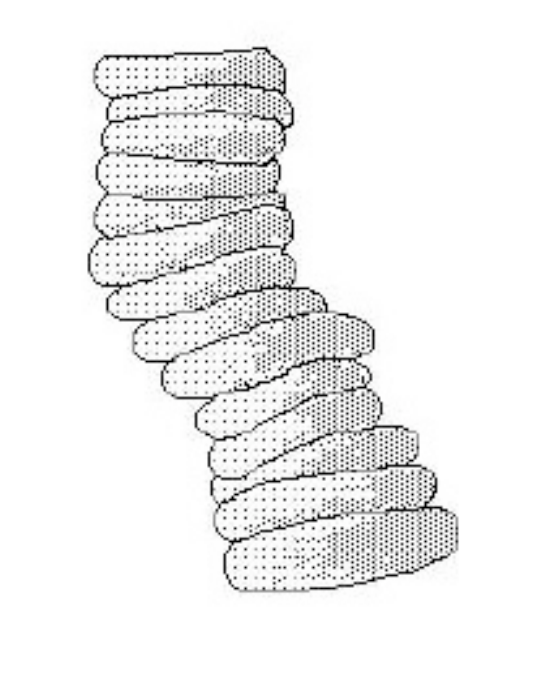}
\caption{\label{Fig0}Sketch of rouleaux formation by erythrocytes.}
\end{minipage}\hspace{2pc}%
\begin{minipage}{12pc}
\includegraphics[width=12pc]{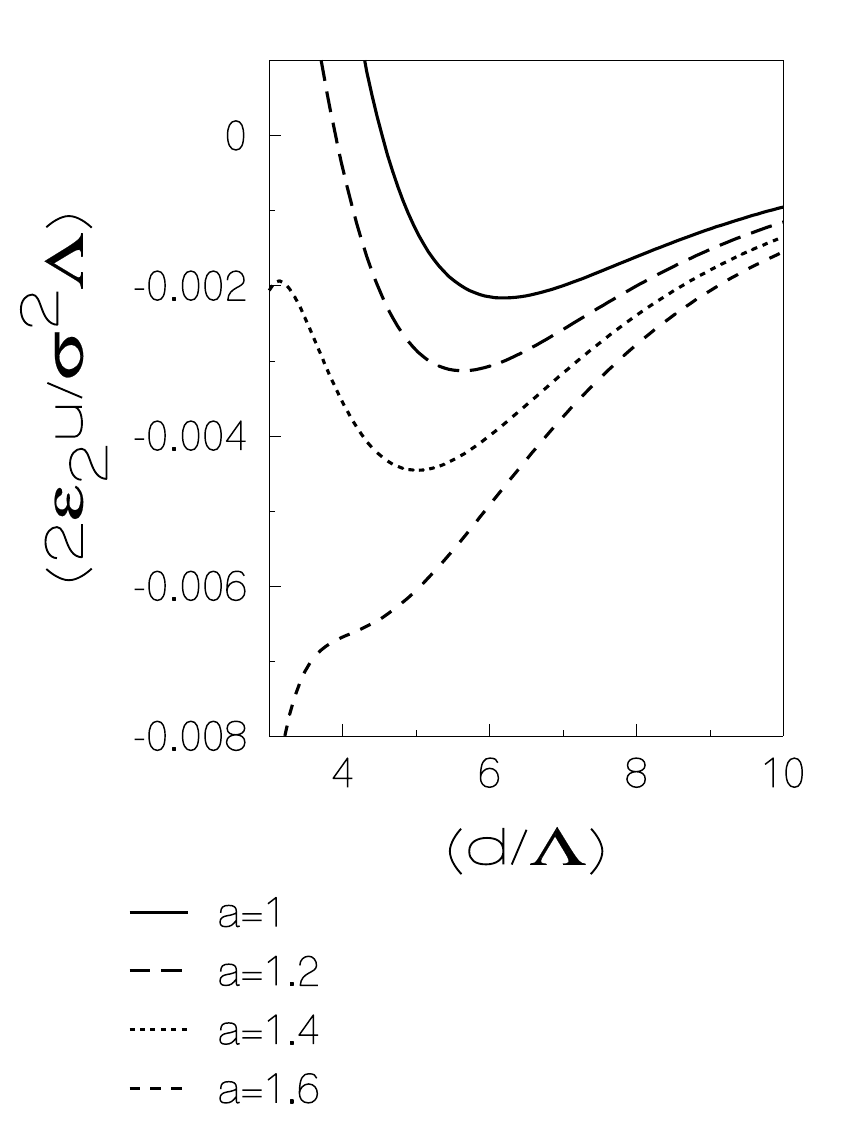}
\caption{\label{Fig1} Free energy per unit area
as a function of separation distance. }
\end{minipage}\hspace{2pc}%
\begin{minipage}{12pc}
\includegraphics[width=12pc]{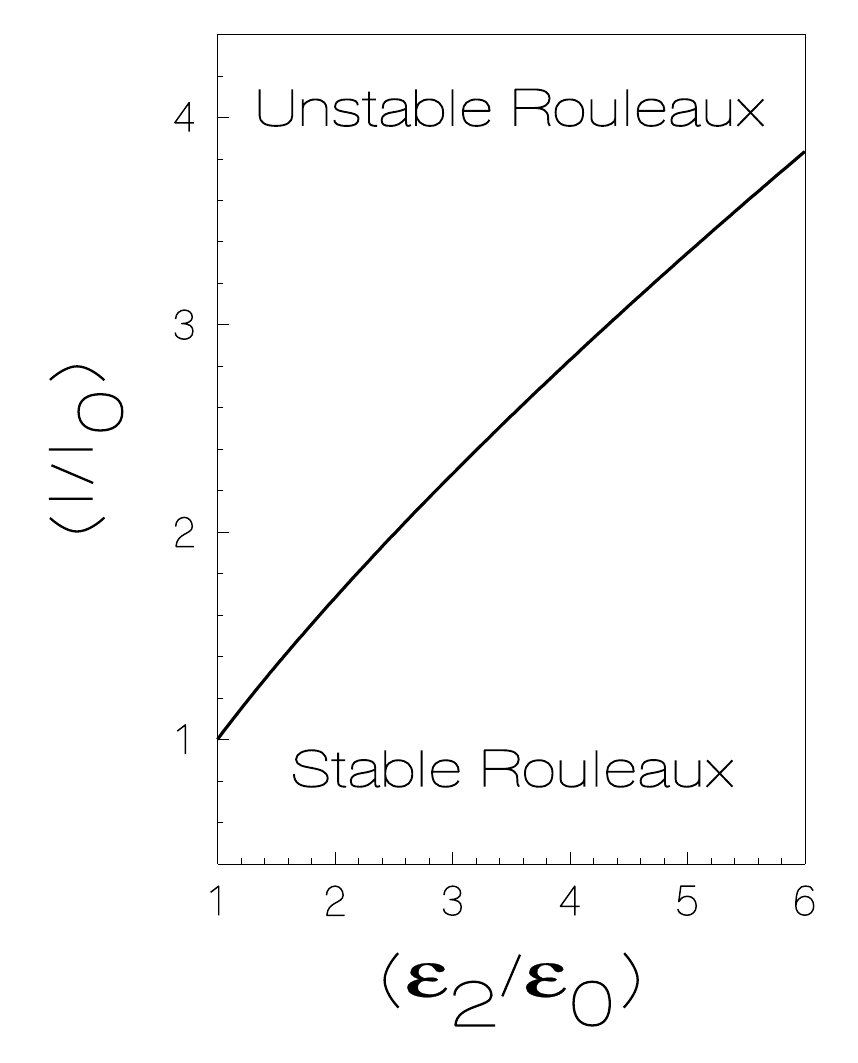}
\caption{\label{Fig2}Phase diagram for
rouleaux formation. }
\end{minipage}
\end{center} 
\end{figure}

At play is also a lateral Casimir force which tends to pull the plates parallel to each other to increase
the negative Casimir contribution to the total energy, which increases with degree to which the plates
overlap and face one another. In this case not only does the Casimir force maintain the stability of
rouleaux, but in fact drives their assembly. Such effects could form the basis
not only of simpler ways to study the Casimir effect (present approaches use external mechanical
forces to control plate separation), but also a variety of nanofabrication \cite{Drexler} techniques
with possible geometries involved not needing to be that of plates. Additional external
fields can be considered such as temperature or external electric fields.

Colloidal aggregation via Casimir forces has also been seen in a non-blood system\cite{Casimir-aggreg}.

\section{Conclusions}

Material properties can be radically altered or materials bound or assembled by external classical
fields or by modification of the quantum fields in the vacuum. Two examples are given in some detail:
the floating water bridge in which liquid water is maintained under tension (negative pressure) by
an external electric field, and the assembly and binding of rouleaux of erythrocytes by the quantum
vacuum forces via the Casimir effect (which also provides a negative pressure). The field is very much
open for further investigations.

\section{Acknowledgements}

The authors would like to thank the conference organizers for a very enjoyable meeting. J. S. is 
partially funded by a grant from the US National Science Foundation. He would also like to thank
M. Babaei for interesting discussions.

\end{document}